\documentclass[aip,jcp]{revtex4-1}
\usepackage{graphicx}

\begin{document}


\title{Studies on electrostatic interactions of colloidal particles under two-dimensional confinement}



\author{Chi-Lun Lee}
\email{chilun@ncu.edu.tw}
\author{Sio-Kit Ng}
\affiliation{Department of Physics and Institute of Biophysics, National Central University, Jhongli 32001, Taiwan}

\date{\today}

\begin{abstract}
We study the effective electrostatic interactions between a pair of charged colloidal particles without salt ions while the system is confined in two dimensions.  In particular we use a simplified model to elucidate the effects of rotational fluctuations in counterion distribution.  The results exhibit effective colloidal attractions under appropriate conditions. Meanwhile, long-range repulsions persist over most of our studied cases. The repulsive forces arise from the fact that in two dimensions the charged colloids cannot be perfectly screened by counterions, as the residual quadrupole moments contribute to the repulsions at longer range. And by applying multiple expansions we find that the attractive forces observed at short range are mainly contributed from electrostatic interactions among higher-order electric moments. We argue that the scenario for attractive interactions discussed in this work is applicable to systems of charged nanoparticles or colloidal solutions with macroions.
\end{abstract}


\maketitle

\section{Introduction}
The interactions among colloidal particles have been extensively investigated for over a century. Although most of the facts about colloidal interactions can be understood following the success of the DLVO theory\cite{DLVO, Israelachvili}, recent observations have revealed striking results that cannot be fully explained yet. In particular, the DLVO theory predicts an effective screened-Coulomb repulsion and a van der Waals attractive force among colloidal particles, as the attractive force exists at pretty short, contact range. However, the so-called ``like-charge attractions'' were observed for colloids confined between glass plates\cite{Fraden94, Tinoco96, Grier96, Grier03} and colloids at aqueous interfaces\cite{Peng05, Peng06}, for metastable colloidal crystallites\cite{Grier97}, and for inorganic macroionic solutions\cite{Liu04, Liu05}. The attractions observed in these experiments cannot be accounted by the van der Waals interactions, as these attractions exhibit at a relatively longer range compared with the size for colloids. For colloidal particles at aqueous interfaces, the mechanism of like-charge attractions was attributed to the lateral asymmetry of counterion clouds, which gives rise to a net horizontal dipole moment for each colloid\cite{Peng05, Peng06}. On the other hand, the mechanisms for like-charge attractions in most other systems are less clear, as it is generally believed that such a relatively long-range interaction must arise from electrostatic interactions. It is also intriguing to notice that the phenomena of like-charge interactions have also been observed for polyelectrolytes\cite{Wong03_1, Wong03_2} and biopolymers such as DNA and F-actin filament\cite{Bloomfield96_1, Bloomfield96_2, Tang96}. In these observations, like-charge attractions among macromolecules were found with the addition of multivalent salt, as the correlated charge fluctuations of counterions were observed\cite{Wong03_1} and regarded as the major source of like-charge attractions\cite{Liu97}.

Theoretically, the sources of like-charge attractions have been accounted by Sogami and Ise using a generalized Poisson-Boltzmann approach\cite{Sogami84}. Later on some non-mean-field theory was developed\cite{Tinoco98, Tinoco02} following a self-consistent Ornstein-Zernike approach proposed by Zerah and Hansen\cite{Hansen86}. However, it is worth noting that in these approaches the isotropic condition is assumed, as the angular fluctuations among counterions are suppressed therein. Alternatively, the effects of counterion polarization have been studied\cite{Levin99, Levin02}, as the results fail to exhibit effective attractions for colloidal separations larger than the Debye screening length.

In order to study the contributions from the angular correlations of counterions and to find further sources of like-charge interactions, we apply in this work a simplified two-dimensional model such that for each colloidal particle, its associated counterions are restricted to move on a ring centered at the colloid. By this setup we suppress the radial fluctuations of counterions and focus on the angular fluctuations instead.

The rest of the paper is organized as follows. In the next section, we use a simple scenario to show that in a two-dimensional colloidal system, the colloidal particles are subjected to effective repulsions which are stronger than typical screened-Coulomb interactions. The framework of our model is given in Sec.~\ref{Model}, as our results in Sec.~\ref{Results} show that in addition to long-range repulsions, effective attractions do demonstrate under certain conditions. The mechanisms for repulsions and attractions are further studied in Sec.~\ref{Simple_model} following a simplified model, which excludes the formation of induced dipoles. Finally, the concluding remarks are given in sec.~\ref{Conclusion}.

\section{The absence of perfect charge screening in two dimensions}
\label{2d_repulsion}
\begin{figure}
  \includegraphics[width = 10cm, height = 8cm]{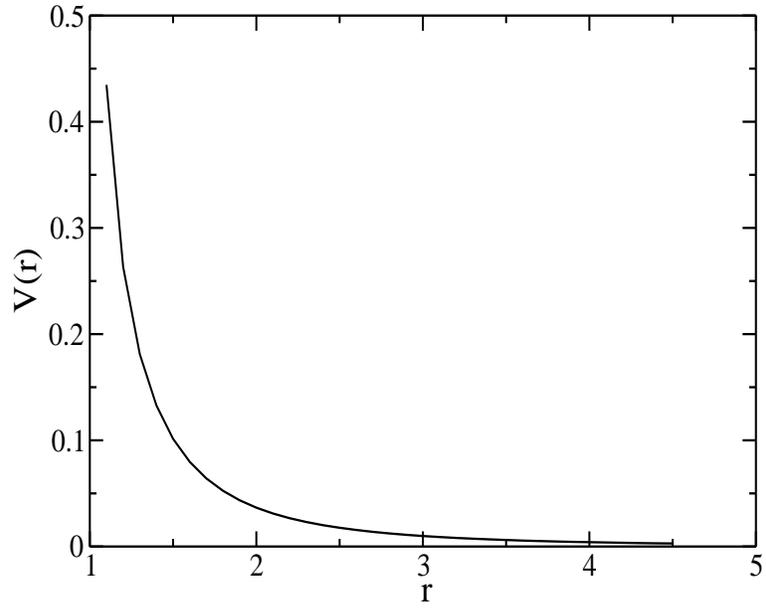}
  \caption{Electrostatic potential versus distance for a negative point charge along with a compensating positively-charged ring. For simplicity we set $q=1$ and $R=1$ (all quantities concerned are represented in dimensionless units).}
  \label{Elliptic}
\end{figure}

\begin{figure}
  \includegraphics[width = 10cm, height = 8cm]{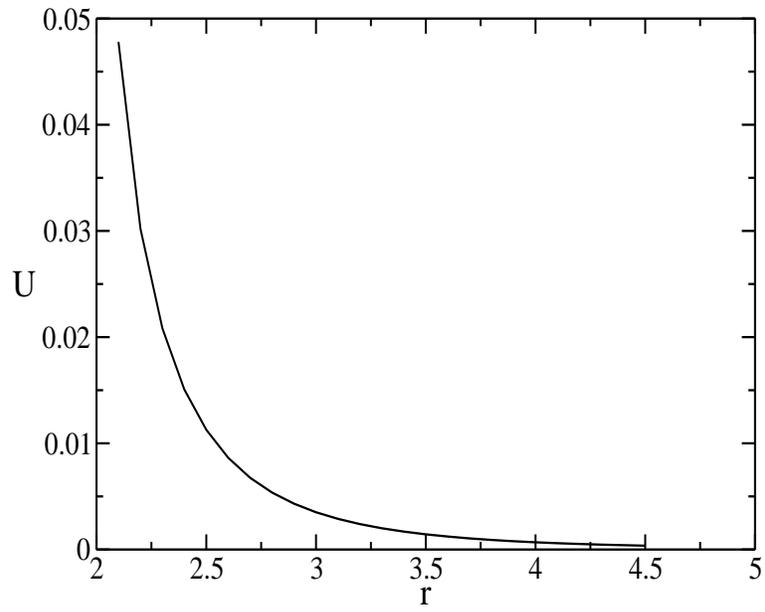}
  \caption{Electrostatic energy versus distance between two (point charge + ring) systems (all quantities concerned are represented in dimensionless units).}
  \label{U_monopole}
\end{figure}

An important feature for two-dimensional charge systems is that perfect charge screening does not exist. To prove this property, let us consider a single negative point charge $-q$ surrounded by a uniformly-distributed positive charge ring with linear density $\lambda= q/(2\pi R)$, where $R$ is the distance to the central negative charge. The three-dimensional analogy of such a setup results in a perfect screening outside the positively-charged spherical shell. However, in two dimensions the charged ring over-compensates the electric field of the point charge $q$, as the electric potential of this system is given by
\begin{eqnarray}
  V(r) &=& -\frac{q}{\epsilon r} + \int_0^{2\pi} \frac{\lambda R\, d\phi}{\epsilon \sqrt{R^2+r^2-2Rr\cos \phi}} \nonumber \\
  &=& -\frac{q}{\epsilon r} + \frac{\lambda}{\epsilon} K\left(2\sqrt{\frac{rR}{(r+R)^2}}\right) \, ,
\end{eqnarray}
where $r$ is the distance to the point charge $q$, and $K(r)$ is the complete elliptic integral of the first kind. By plotting $V(r)$ in Fig.~\ref{Elliptic} for the example $q=1$, $\epsilon =1$, and $R=1$ (if all quantities are represented in dimensionless units), we find that this residual potential is positive, indicating that the electric field of the central charge is over compensated by the surrounding positive charge ring. Also we plot the electrostatic energy between a pair of such (point charge + ring) systems in Fig.\ref{U_monopole} versus the distance, as the result exhibits repulsions at all range.

Thus one can treat the colloidal system as an effective soft-sphere solution, while the effective diameter is determined by the counterion distribution instead of the colloidal size. In a dilute suspension of colloidal particles, the effective diameter is much larger than that for the colloidal particles. Besides this repulsive nature for colloids in two dimensions, it is tempting to explore the possible mechanisms in electrostatics and thermostatistics that lead to an effective attractive force between a pair of colloidal particles. In the next section we use a simplified model to study the possible sources of attractions.

\section{Model}
\label{Model}
Contrary to the framework of most theoretical studies in colloidal systems, which lay attention on the radial fluctuations of counterions about colloidal particles, we choose to emphasize the importance of the rotational fluctuations instead. And in this study we decide to leave alone the van der Waals interactions among colloidal particles, as we aim to study the electrostatic interactions of charged colloids along with their counterions.

\begin{figure}
  \includegraphics[width = 10cm, height = 4.4cm]{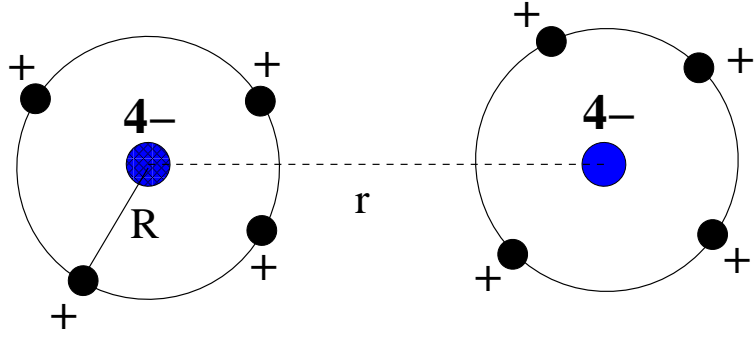}
  \caption{Illustration for our model. In this model the counterions for each colloidal particle are restricted such that they can only move on a ring. This facilitates our studies for the effects of rotational fluctuations of counterions.}
  \label{illu1}
\end{figure}

We set in our model that each colloidal particle has a net charge $-nq$ located at its center.  There are $n$ surrounding counterions, each bearing a charge $+q$.  Since we are interested in the rotational fluctuations of these counterions, we make the restriction such that these counterions are distributed on a ring about the colloidal particle, and $R$ is the radius of the ring.  (Please refer to Fig.~\ref{illu1} for an illustrative example.)  Assuming that there is no added salt, the total electrostatic energy among a pair of colloidal particles and their counterions can be written as
\begin{equation}
  \frac{U^0}{k_\mathrm{B} T} = \sum_{\mbox{all\ charges}} \frac{q_i q_j}{\epsilon r^0_{ij} k_\mathrm{B} T} = \sum_{\mbox{all\ charges}} \frac{z_i z_j}{r_{ij} t}\, .
\end{equation}
where $z_i = q_i/q$ is the valence of the $i$th charge, $\epsilon$ is the dielectric constant, and $r^0_{ij}$ is the distance between the $i$th and $j$th charge.  In this work we use dimensionless variables such that $r_{ij}\equiv r^0_{ij}/R$, and we introduce a reduced temperature $t$ such that $t \equiv \epsilon k_\mathrm{B} T R/q^2$.  Also the dimensionless analog of the electrostatic energy is defined as $U \equiv U^0 t /(k_\mathrm{B} T)$. Note that $1/t$ is proportional to the Bjerrum length $\ell_\mathrm{B}$ in units of $R$.  To derive effective potentials and correlations, we perform Monte Carlo simulations via the Metropolis algorithm. .

\section{Results and discussions}
\begin{figure}
  \includegraphics[width = 8cm, height = 20cm]{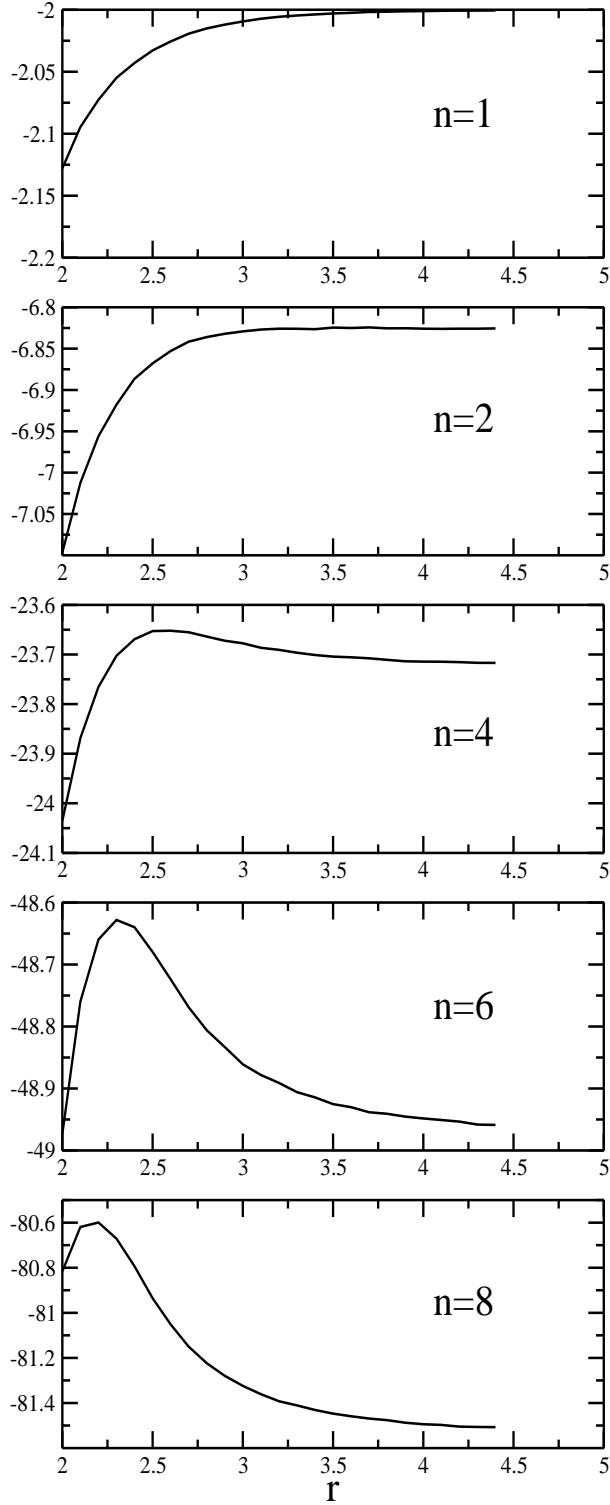}
  \caption{Average electrostatic potential energy versus distance (in dimensionless units) between two colloidal particles $r$ at the reduced temperature $t=0.25$ for various number of
counterions $n=1$, 2, 4, 6 and 8.}
  \label{U_r_T0_25}
\end{figure}

\begin{figure}
  \includegraphics[width = 10cm, height = 14cm]{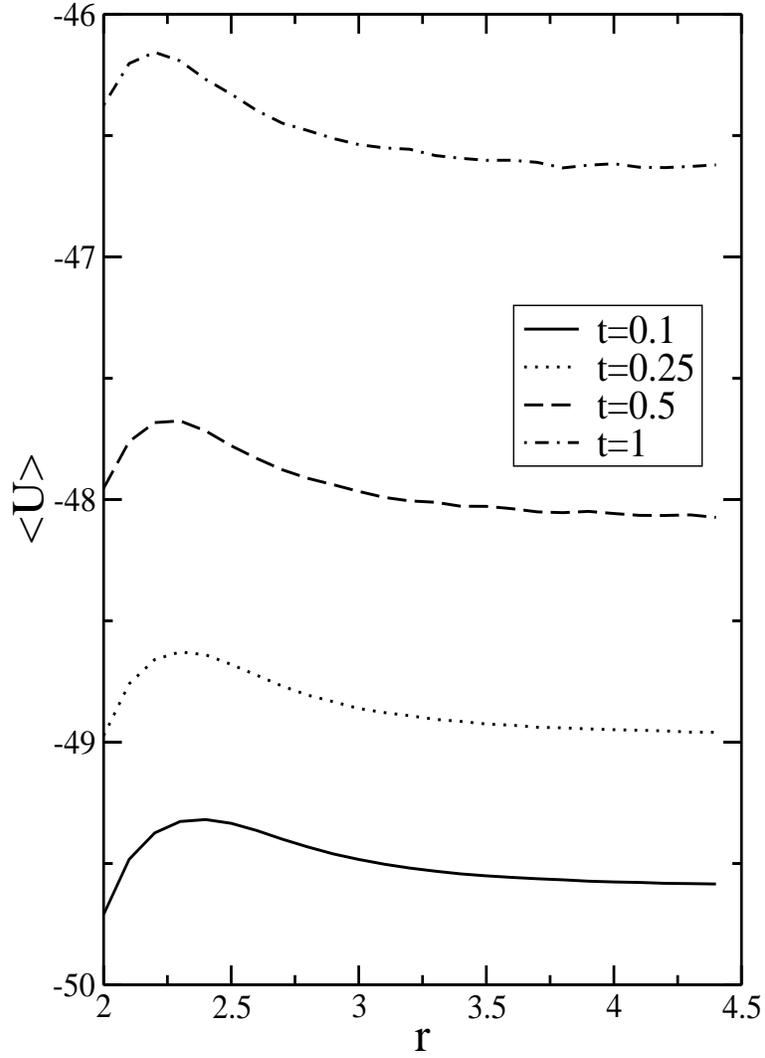}
  \caption{Average electrostatic potential energy versus distance (in dimensionless units) between two colloidal particles $r$ at reduced temperatures $t=0.1$, 0.25, 0.5 and 1 for $n=6$.}
  \label{U_r_n6}
\end{figure}

]
\label{Results}
Fig.~\ref{U_r_T0_25} shows the average dimensionless electrostatic energy $U$ of a two-colloidal system for colloidal charges $n=$1, 2, 4, 6, and 8, as $r$ is the distance between the centers of the two colloidal particles.  We find that for $n=1$ and 2 the average potential energy profile is mostly attractive, while for $n=$4, 6, and 8 the potential clearly exhibits a short-range attractive part as well as a repulsive part at longer range. From these results the depth of short-range attractions is comparable to the magnitude of thermal energy, and the range of attractions diminishes as the number of counterions $n$ increases.  For example, at $n=4$ the turning point of the average potential is $r\approx 2.6$, while the depth of the attractive potential is $\Delta U \approx 0.38$.  The corresponding numbers are $r\approx 2.2$ and $\Delta U\approx 0.21$ for $n=8$.  Moreover, the repulsive part of the potential emerges and becomes more dominant as $n$ increases. This is because that for larger $n$ the counterions are more uniformly distributed on the ring due to their mutual repulsions. And this close-to-uniform distribution results in a scenario that is similar to what we have discussed in Sec.~\ref{2d_repulsion}. In Fig.~\ref{U_r_n6} the electrostatic energy is shown for the case $n=6$ at various temperatures. The result shows that the attractive range diminishes while the temperature increases, and the repulsive tail persists at all temperatures considered.

From the results above we find that effective colloid-colloid attractions can arise for the case of small $n$. The effective energy also exhibits a repulsive tail outside the counterion shell, which is a signature for colloidal systems in two dimensions. The striking result that both attractions and repulsions are observed in the effective potential is reminiscent of the DLVO theory. However, the mechanisms for attractions and repulsions in our model are quite different from those accounted in the DLVO theory. For example, we do not consider in our model the van der Waals interactions among colloidal particles, as all the features we observe must be attributed to electrostatic interactions only.

\begin{figure}
  \includegraphics[width = 10cm, height = 8cm]{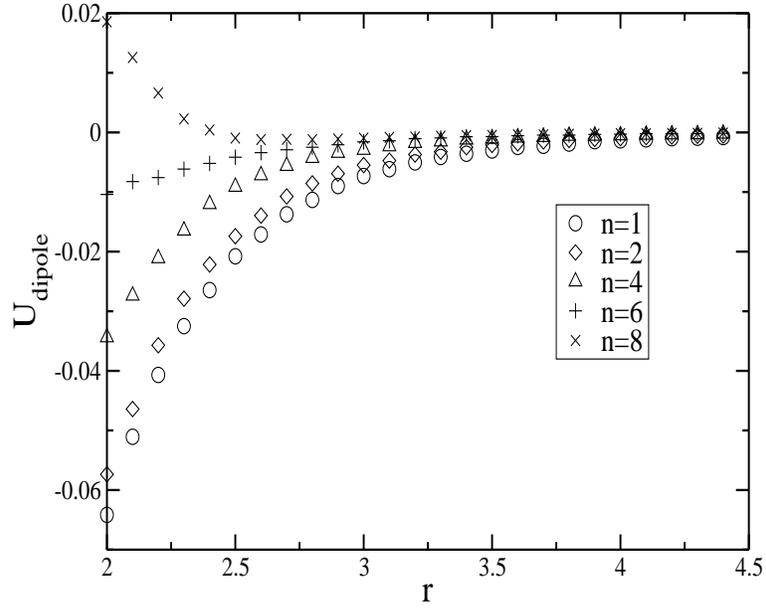}
  \caption{Average dipole-dipole energy versus distance $r$ (in dimensionless units) at reduced temperatures $t=0.25$ for $n=1$, 2, 4, 6 and 8.}
  \label{U_dipole}
\end{figure}

To find the sources of attractive interactions, we first look at the induced dipole-dipole interactions. From our Monte Carlo simulations we record the contributions of the dipole-dipole interactions via the following average:
\begin{equation}
  U_{\mathrm{dipole}} = \left\langle \frac{\mathbf{p}_1 \cdot \mathbf{p}_2 - 3 (\mathbf{p}_1\cdot \hat{r})(\mathbf{p}_2\cdot \hat{r})}{r^3} \right\rangle \, ,
  \label{eq_U_dipole}
\end{equation}
while $\mathbf{p}_1$ and $\mathbf{p}_2$ are the dipole moments of the two colloids with their counterions, respectively, and $\hat{r}$ is the unit vector aligned with the centers of the two colloids. The result of $U_{\mathrm{dipole}}$ at $t=0.25$ is shown in Fig.~\ref{U_dipole}. One finds that for all cases the magnitude of the dipole-dipole interactions is too small to account for the attractive interactions shown in Fig.~\ref{U_r_T0_25}. For small $n$ such as $n=1$, the actual effect of dipole-dipole interactions must be enlarged knowing that in Eq.~\ref{eq_U_dipole} we actually compute the electrostatic energy between point dipoles. On the other hand, our system of colloids consists of physical dipoles with a size comparable to the distance between colloids. However, this still does not suffice to account for the attractions observed in Fig.~\ref{U_r_T0_25}, because the polarizability is greatly reduced as $n$ increases. For example, for $n=4$ and $6$ the magnitude of $U_\mathrm{dipole}$ turns out to be one order of magnitude smaller than the attractive part in the overall effective energy. Moreover, for the case $n=8$ one finds that $U_\mathrm{dipole}$ is mainly repulsive, while the overall electrostatic energy still exhibits an attractive part at the short range. Therefore we need to find further sources of electrostatic interactions that result in the short-range attractions.

\begin{figure}
  \includegraphics[width = 10cm, height = 8cm]{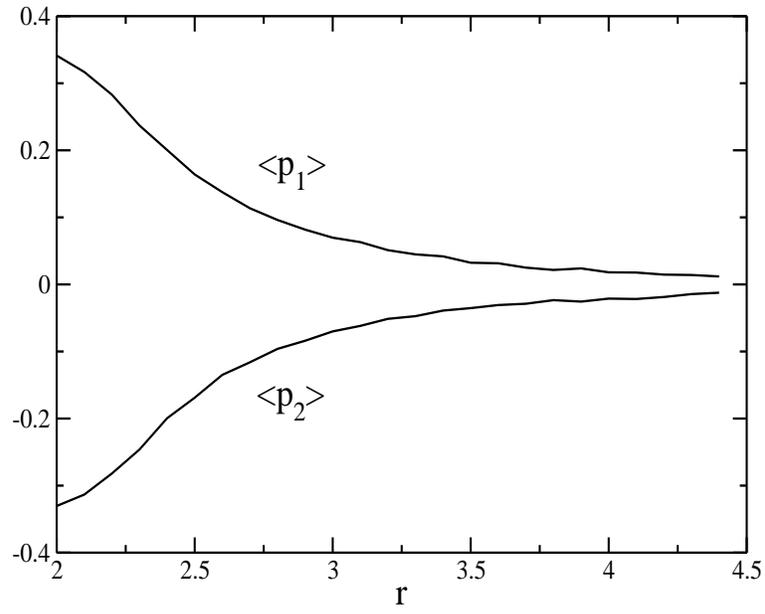}
  \caption{Average dipole moments versus distance $r$ (in dimensionless units) at reduced temperatures $t=0.25$ for $n=6$.}
  \label{p_r_n6}
\end{figure}

In Fig.~\ref{p_r_n6}, the average dipole moments of the two colloids along with their counterions are presented for the case $n=6$ at $t=0.25$. Due to symmetry the two averaged dipole moments must have opposite directions. Nevertheless, the nonvanishing average dipole moments imply that after average over thermal fluctuations, a colloidal particle with its counterions still result in a residual electric field even outside the counterion shell, which in turn polarizes the counterions of the other colloidal particle. From the direction of the average dipole moment one learns that the net electric field of each colloidal particle along with its counterions points out from its center, showing an evidence that in two dimensions the counterions cannot perfectly cancel out the electric field of the charged colloidal particle, as the residual compensation causes an electric field that is radiating out of the ring. And this residual electric field causes mutual repulsions between colloidal particles at larger distance, which we have already discussed above in Sec.~\ref{2d_repulsion}, and we will be further investigate this feature via the multipole expansion method in the following section.

\section{Sources of attractive and repulsive interactions}
\label{Simple_model}
\begin{figure}
  \includegraphics[width = 10cm, height = 8cm]{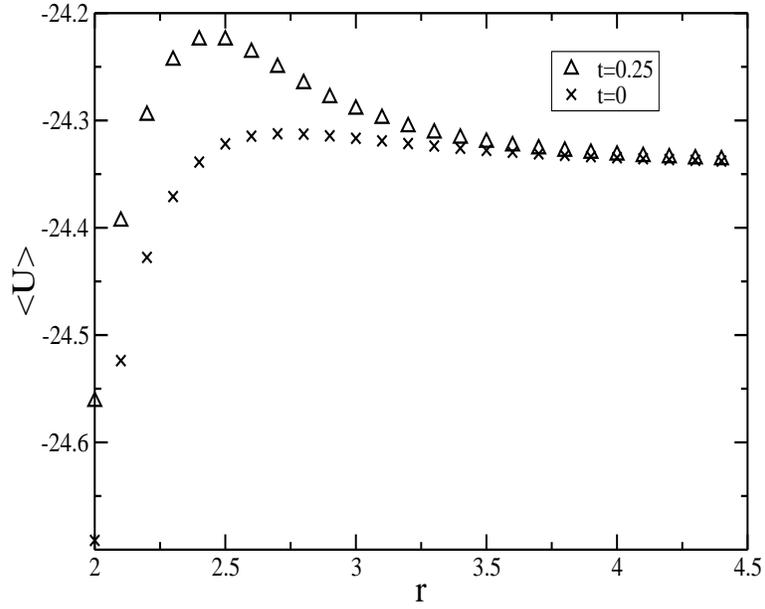}
  \caption{Average electrostatic potential energy versus distance (in dimensionless units) at reduced temperatures $t=0.25$ for $n=4$, with the restriction that the counterions are evenly distributed about the colloids.}
  \label{num4_fixed}
\end{figure}

To study the sources of attractions other than the induced dipole-dipole interactions, we use a test model such that the counterions are evenly distributed about the center of the corresponding colloidal particle. With such a strong restriction the counterions can only rotate collectively about the colloidal particle during the course of simulations, and for this test model there are no induced dipoles due to symmetry considerations.  For the case $n=4$ the result at $t=0.25$ is shown in Fig.~\ref{num4_fixed}. Again we observe both the attractive and repulsive parts in the average potential energy, as the depth of the attractive potential is of the same magnitude compared with our original model (see Fig.~\ref{U_r_T0_25}). Furthermore, Fig.~\ref{num4_fixed} shows that the attractive part persists even in the zero-temperature limit, while the repulsive part weakens greatly.

\begin{figure}
  \includegraphics[width = 10cm, height = 4.1cm]{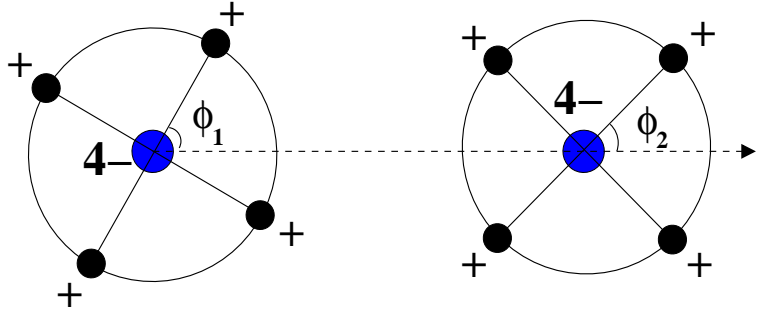}
  \caption{Illustration for our simplified model . We use this test model such that the counterions are evenly distributed about the center of the corresponding colloidal particle. The counterions can only rotate collectively about the colloidal particles, as we use $\phi_1$ and $\phi_2$ to characterize their positions.}
  \label{illu2}
\end{figure}

To understand the origin of the attractive and repulsive interactions, we perform a series expansion of the electrostatic energy with respect to the inverse distance $1/r$. In this simplified model each configuration can be characterized through two angles $\phi_1$ and $\phi_2$ (as shown in Fig.~\ref{illu2}). By the series expansion on gets
\begin{equation}
U(r,\phi_1,\phi_2) = \frac{A}{r^5}+\frac{B}{r^7}+\frac{C}{r^9}+O\left(\frac{1}{r^{11}}\right) \, .
\label{multipole_exp}
\end{equation}
For the case $n=4$, one has
\begin{eqnarray}
  A&=&9\, ,\nonumber \\
  B &=& \frac{45}{16} \left[ 10 + 7\cos(4\phi_1) + 7\cos(4\phi_2)\right]  \, , \ \mbox{and} \nonumber \\
  C &=& \frac{7}{512} \left[ 5950 + 5346 \cos(4\phi_1) + 5346 \cos(4\phi_2) \right.  \nonumber \\
    && \ \ \ \ \left. + 175 \cos(4\phi_1 - 4\phi_2) + 32175 \cos(4\phi_1 + 4 \phi_2)\right] \, .
  \label{multi_exp_coeff}
\end{eqnarray}
Since our simplified model allows no dipole moment for counterion distributions on each colloids, the lowest order in the expansion starts from the $1/r^5$ term, which represents quadrupole-quadrupole interactions. The independence of $A$ on $\phi_1$ and $\phi_2$ reflects the fact that the quadrupole moment is constant regardless of the overall counterion orientation. This quadrupole-quadrupole interaction is repulsive and is dominant at large distances.

For this simplified model one finds that due to symmetry there is no octupole moment, as the next nonvanishing term in the expansion Eq.~\ref{multipole_exp}, the $O(1/r^7)$ term, is attributed to the hexadecapole-quadrupole interactions. In Table~\ref{table} we list the contributions of the first few terms in the lowest-energy state. One finds that although the $O(1/r^7)$ term could be attractive with suitable choices of $\phi_1$ and $\phi_2$, the effect is overpowered by the higher-order terms (such as the $O(1/r^{9}$ term) at short range. Therefore, the origin of short-range attractions in the zero-temperature limit is contributed from the hexadecapole-hexadecapole ($O(1/r^9)$) and higher-order multipole interactions.

\begin{figure}
  \includegraphics[width = 10cm, height = 8cm]{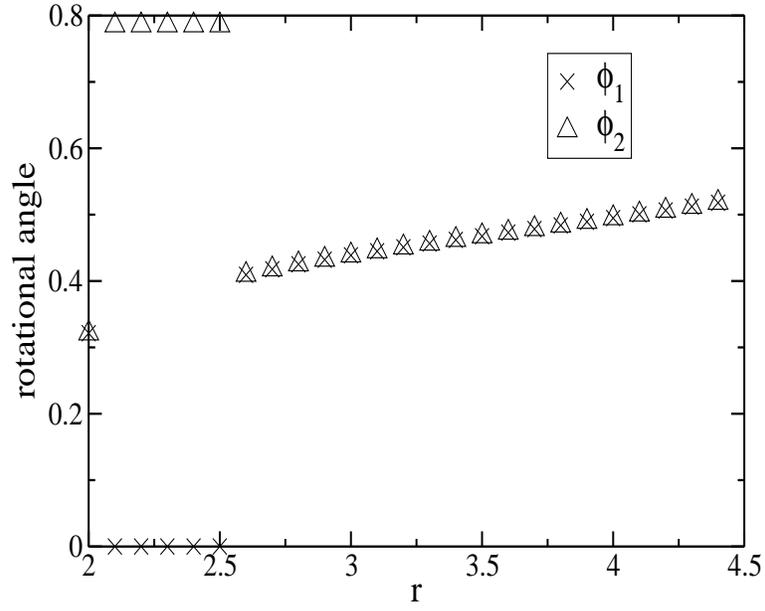}
  \caption{Rotational angles $\phi_1$ and $\phi_2$ of counterions in the zero-temperature limit for $n=4$, with the restriction that the counterions are evenly distributed about the colloids.}
  \label{angle_T0_num4}
\end{figure}

\begin{figure}
  \includegraphics[width = 10cm, height = 8cm]{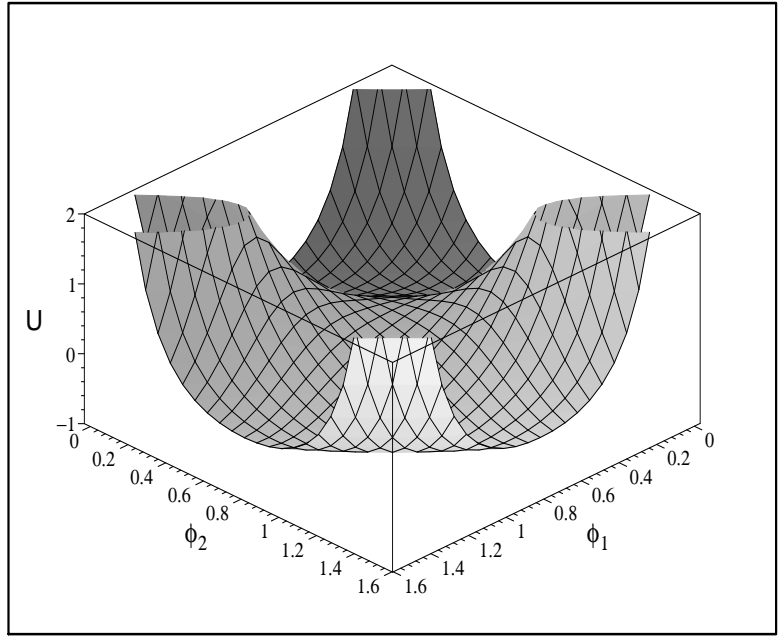}
  \caption{The three-dimensional energy landscape for the simplified model at $r=2.1$ and $n=4$ ($U$ and $r$ are defined in dimensionless units). }
  \label{U_T0_3D}
\end{figure}

In Fig.~\ref{angle_T0_num4} we list the counterion rotational angles $\phi_1$ and $\phi_2$ in the zero-temperature limit. We observe that there exists discontinuous transitions at $r\approx 2.1$ and $r\approx 2.6$. This result implies the existence of local minima in the energy landscape, and the lowest-energy state is shifting from one to another local minima at these transitions. Alternatively, these transitions can also be realized by the fact that the various terms in Eq.~\ref{multipole_exp} are competing from one another. As the distance $r$ increases, the relative weights of these contributions are also modulated, and the consequence can be the emergence of a new local minimum, the disappearance of some preexisting local minimum, or just the switching of relative order between two local minima. In Fig.~\ref{U_T0_3D} we show the energy landscape over $\phi_1$ and $\phi_2$ at $r=2.1$. We find that although the energy minimum lies at $(\phi_1=0$, $\phi_2=\pi /4)$ (as well as $(\phi_1=\pi /4$, $\phi_2=0)$), there exists valleys on the energy landscape of which the energy is very close to the global minimum, as the deviation is less than 0.01. This feature implies that correlated fluctuations would emerge even in the low-temperature regime (such as $t\approx 0.01$).

From the results above, we can conclude that the repulsive interactions between colloidal particles arise mainly from the quadrupole-quadrupole interactions, which is itself a special feature in two-dimensional systems. In fact if one computes the pair electrostatic interactions for the scenario in Sec.~\ref{2d_repulsion} and makes a series expansion over $1/r$, he can find that the leading term starts from the quadrupole-quadrupole interactions ($O(1/r^5)$) as well. The attractive interactions between colloidal particles are due to the correlated interactions among higher-order moments, as in this test model, and for the case $n=4$, the attractive interaction starts from the hexadecapole-hexadecapole ($O(1/r^9)$) interactions.

\section{Conclusion}
\label{Conclusion}
From our studies that focus on the rotational fluctuations of counterions, we observe the effective interactions which contain for most of our cases a repulsive tail and short-range attractions between colloidal particles. The repulsive part of effective interactions is not a characteristic of screened Coulomb interactions between colloidal particles. Instead, this special type of repulsions persists at a longer range compared with the ring diameter (the latter of which often being regarded as the range of screening). This result exhibits a distinct feature for two-dimensional colloidal systems. This feature originates from the fact that a charged colloidal particle cannot be perfectly screened by its counterions embedded in two dimensions. Instead its electric field is over compensated by the counterions, as the residual quadrupole moment causes the effective repulsions.  The effective attractions between colloidal particles can be observed when the number of counterions $n$ is small (in our studies it can be observed up to $n=8$). Our results show that the counterions are distributed in a way such that there exists short-range attractions in the effective interaction energy. The distribution of counterions can be analyzed via multipole expansions, as we find for the case $n=4$ that the attractions arise from the hexadecapole-hexadecapole ($O(1/r^9)$) and higher-order terms.

Quite often effective attractions are observed between a pair of charged systems, each being charge neutral on the whole, if the system allows enough degrees of freedom. For example, a pair of free-rotating electric dipoles have an effective electrostatic energy that is attractive\cite{Israelachvili}. In this work we find just another example exhibiting effective attractions between a pair of colloidal particles along with their counterions, while we consider the rotational degrees of freedom only for the counterions. Meanwhile we have shown that for such systems, the induced multipole moment weighs much more than the induced dipolar contribution in the overall electrostatic interactions. Although our work is done in two dimensions, we have tried to perform some simulations for an analogous model in three dimensions, and the preliminary result also shows effective short-range attractions with similar parameter settings. But contrary to the two-dimensional model, we have not found any significant repulsions at longer range.

The reduced temperature defined in this work is linked to the Bjerrum length via the relation $\ell_\mathrm{B}/R = e^2/(q^2 t)$. If the counterions are monovalent, the range of reduced temperature studied in this work corresponds to a Bjerrum length of the order that can be compared to the counterion distribution radius $R$. This implies that for micrometer-scaled colloidal particles in aqueous solutions, such attractive mechanism can emerge at a normal Bjerrum length only if the counterions are themselves macroions (such that $q$ is large). On the other hand, this type of attractions might be observed for nanometer-sized macroions such as proteins or other inorganic macroions\cite{Liu04, Liu05}. The reason that in our results attractions can be observed for rather small number of counterions may be attributed to the fact that once we lock the counterions on the ring, their mutual repulsions allow for less fluctuations. As there are enough space for counterions, anisotropic fluctuations should emerge, as the subsequent induced multipole-multipole attractions can be observed at a higher reduced temperature.

\begin{acknowledgments}
This work was supported by the National Science Council of the Republic of China under Grant No. NSC-98-2112-M-008-009.
\end{acknowledgments}

\clearpage
\begin{table}
  \begin{tabular}{|c|c|c|c|c|c|}
   \hline
   r & 2.1 & 2.4 & 2.7 & 3.0 & 4.0  \\
   \hline
   $ \delta U(r)$ & -0.181 & $4.35 \times 10^{-3}$ & $3.08 \times 10^{-2}$ & $2.66 \times 10^{-2}$ & $8.45\times 10^{-3}$ \\
   \hline
   $A/r^5$ & 0.220 & 0.113 & $6.27\times 10^{-2}$ & $3.70\times 10^{-2}$ & $8.79\times 10^{-3}$ \\
   \hline
   $B/r^7$ & 0.156 & $6.13\times 10^{-2}$ & $2.31 \times 10^{-2}$ & $9.56\times 10^{-3}$ & $7.62\times 10^{-4}$ \\
   \hline
   $C/r^9$ & -0.454 & -0.137 & $-4.75\times 10^{-2}$ & $-1.80\times 10^{-2}$ & $-1.05\times 10^{-3}$\\
   \hline
  \end{tabular}
  \caption{Numerical results of the electrostatic energy $\delta U(r)  \equiv U_\mathrm{min}(r) - U(r=\infty)$, and contributions from the first few terms in the series expansion. The coefficients $A$, $B$, and $C$ are defined in Eq.~\ref{multi_exp_coeff}. The numbers are represented in dimensionless units.}
  \label{table}
\end{table}

\clearpage

\clearpage

\clearpage

\clearpage

\clearpage

\clearpage

\clearpage

\clearpage

\clearpage

\clearpage

\clearpage

\end{document}